\documentclass[fleqn,usenatbib,useAMS]{mnras}
\usepackage[dvips]{graphicx}
\usepackage{xfrac}
\usepackage{xcolor}
\usepackage{array}
\usepackage{amsmath,amssymb,amsfonts}
\usepackage{latexsym}
\usepackage{epstopdf} 
\usepackage{wasysym}
\usepackage{float}
\usepackage{array,multirow}
\begin{document}
\title[Fluctuations in Nova Light Curves]{Explaining Prolonged Fluctuations in Light Curves of Classical Novae via Modeling}

\author[Y. Hillman]{
	Yael Hillman$^{1}$\thanks{Yael Hillman e-mail: yaelhi@ariel.ac.il}\\
	$^{1}$Department of Physics, Ariel University, Ariel, POB 3, 4070000, Israel\\
}

\date{Accepted XXX. Received YYY; in original form ZZZ}

\pubyear{2022}

\label{firstpage}
	\pagerange{\pageref{firstpage}--\pageref{lastpage}}
\maketitle

\begin{abstract}
Fluctuations during a prolonged maximum have been observed in several nova eruptions, although it is not clear, and can not be deduced directly from observations, if the phenomenon is an actual physical reaction to some mechanism originating in the erupting white dwarf, if it is occurring in the expanding ejected shell or if it is a form of interaction with the red dwarf companion. A handful of erupting nova models are investigated in this work, in order to assess the possibility of this sort of feature being an actual part of the eruption itself. The results explain that the mechanism that may produce these fluctuations is the repeated approach and recession of the convective front from the surface. The efficiency of this mechanism, being dependent on the mass of the WD envelope and the time scale of the nova cycle, favors low mass WDs and long accretion phases.
\end{abstract}


\section{Introduction}
A pre-maximum halt (PMH) is a brief plateau or small dip in a nova light curve before it reaches maximum brightness \cite[]{McLaughlin1960,Hounsell2010,Hounsell2016}. It occurs right after the ignition of the thermonuclear runaway (TNR), when the optically thick envelope is suddenly extremely hot, and has not yet expanded enough to cool. \cite[]{Starrfield1987,Starrfield2016,Hillman2014,Hillman2014a,Eyres2017,Poggiani2018,Chomiuk2021}. Observations of PMHs are rare, obtained only by luck, since they are usually very short lived --- only a fraction of the rise time, which in itself is short \cite[e.g.,][]{Starrfield1972,Prialnik1978,Prialnik1986,Shara1989,Prikov1995,Bode2012}. The rise time is relatively longer for slower developing novae, thus PMHs, in the form of a short plateau, are more commonly found in slow novae, such as V1548 Aql \cite[]{Kato2001,Primak2003}, HR Del \cite[]{Terzan1970,Drechsel1977,Friedjung1992}, V723 Cas \cite[]{Hirosawa1995,Ohsima1996,Iijima1998} and V5558 Sgr \cite[]{Munari2007,Naito2007,Poggiani2008,Tanaka2011}. A PMH was also observed in the fast nova, V463 Sct \cite[]{Kato2002,Poggiani2018}, which is rarely observed.
Some novae exhibit a PMH in the form of a small amplitude peak/dip in the light curve (typically of order less than one magnitude) and short timescale (typically of order hours) during the rise to maximum, such as seen in RS Oph, KT Eri and V5589 Sgr \cite[]{Itagaki2009,Hounsell2010,Eyres2017}. The nova M31 2009-10b, being observed in the NUV as well as the R-band, showed a halt of about ten days accompanied by a brief peak in the NUV \cite[]{Cao2012} supporting the theory  that these features are the result of expansion lagging behind the sudden heating from the TNR, causing a shift in the black body peak band.  

There are cases, usually in slow nova, where the light curve exhibits, in addition to a PMH, fluctuations of order 1-3 magnitude during a prolonged maximum before showing any sign of decline. Examples of such systems that are rich in recent observations are V612 Sct (= ASASSN-17hx = Nova Sct 2017) \cite[]{Poggiani2018} and even more recently, V1391 Cas \cite[]{Dubovsky2021}.

V612 Sct was discovered in late June 2013 before reaching maximum brightness \cite[]{Saito2017,Stanek2017}. The brightness in the visual band halted (i.e., formed a PMH) and then continued to rise until peaking a few weeks after first detection. This peak soon declined by about three magnitudes, only to be followed by a series of multiple such peaks over the next few months \cite[]{Poggiani2018}. 

V1391 Cas exhibited the same peculiar behavior as V612 Sct. First detected during the rise in July 2020 \cite[]{Sokolovsky2020_13903}, followup observations detecting oscillations \cite[]{Sokolovsky2020_13904,Sokolovsky2020_14004} and then extensively observed over roughly the following six months \cite[]{Dubovsky2021}. During this time the nova produced multiple peaks with amplitudes of about 2-3 magnitudes that lasted a few days each \cite[]{Dubovsky2021,Fujii2021}. \cite{Schmidt2021} have calculated a remarkably accurate periodicity of $3.8036\pm0.0005$ hours over $\sim100$ days of observation. Attributing this to the orbital period, implies that the donor star is of order $\sim0.3-0.4M_\odot$ \cite[]{Knigge2011a,Kalomeni2016,Hillman2020,Hillman2021}. Mass lost from a system due to a nova eruption causes angular momentum loss that results in an increase in the binary separation, meaning that the new accretion phase begins with a relatively lower accretion rate than before eruption \cite[]{Hillman2020,Hillman2021}. The dominant mechanism responsible for removing angular momentum from the system during the long accretion phases is magnetic braking. This loss of angular momentum causes the stellar separation to slowly shrink which causes the accretion rate to slowly increase \cite[]{Ritter1988,Livio1994b,Kolb2001,Hillman2020,Hillman2021}. 
However, stars with masses of less than $\sim0.35M_\odot$ become fully convective \cite[]{Howell2001} rendering the magnetic braking inefficient. Left with only gravitational radiation (which is about one tenth as efficient as magnetic braking) to pull the stars together, the accretion rate increases at a much slower pace. The result of this whole ordeal is that the average accretion rate for systems with fully convective stars is of order $10^{-11}M_\odot yr^{-1}$ --- about 1\% of the average accretion rate in systems with efficient magnetic braking. 
However, since both the magnetic braking turn-off mass and the donor mass of V1391 Cas are both estimates, and roughly the same, the donor mass in V1391 Cas is not indicative of whether or not magnetic braking is active in this system. This means that the estimated mass does not contribute to the constraining of the average accretion rate, which could be typical of a fully-convective or a non-fully-convective star (or anything in between). 

Identifying the main parameters of systems that produce features such as seen in V1391 Cas and V612 Sct (and other slow nova), is the first step in understanding what causes such peculiar features, and why they form in some systems, and not in others. A few theoretical mechanisms have been proposed (but yet to have been proven) to explain these fluctuations,  including (1) disk instabilities, on the account that it has managed to somehow survive the eruption \cite[]{Dubovsky2021}; (2) a rapidly or drastically changing rate of mass transfer, \cite[]{Dubovsky2021} which should not be erratic in CVs; (3) shocks from the interacting ejecta and donor envelope, provided the density of the donor envelope is inconsistent \cite[]{Aydi2020} and (4) instabilities in the WD envelope that lead to short, multiple ejection episodes \cite[]{Dubovsky2021}. 
This paper explores the latter possibility via modeling of the nova cycle (accretion and eruption)
without the need for interaction with anything exterior to WD. The focus is on the two systems, V1391 Cas and V612 Sct, for convenience and for aiding in visual comparisons, but the conclusions apply to other similar systems as well. 

The following section describes the models, followed by results in \S\ref{sec:results}. The results are discussed in \S\ref{sec:discussion} and the main conclusions are summed in \S\ref{sec:conclusions}.

\section{Models}\label{sec:models}
All but one of the models were produced using a hydrodynamic Lagrangian nova evolution code \cite[]{Prikov1995,Epelstain2007} described in further detail in \cite{Hillman2015}. The basic input parameters for the code are the WD mass ($\rm M_{WD}$) and a constant rate at which it accretes mass from its donor ($\dot{M}$). One model was produced using the combined, self consistent code, which utilizes the aforementioned nova code and a stellar evolution code, described in detail in \cite{Kovetz2009}. The combined code is described in \cite{Hillman2020}.
The requirements for choosing the parameters for a model were for it to resemble the general features and time scales of the first few months after detection of the two novae V612 Sct and V1391 Cas. As detailed above, the observations of both of these nova exhibit (1) an initial rise to maximum on the time scale of order tens of days, implying a low mass WD (see \S\ref{sec:discussion} for elaboration) and (2) a long lasting maximum, of order one hundred days, implying a relatively low rate of mass accretion, typical of classical novae, which can allow this long slow eruption. The additional fascinating feature seen in these two nova (and more, to be discussed in \S\ref{sec:discussion}) are multiple peaks during an otherwise long flat plateau at maximum \cite[see type "O" in][]{Strope2010} which this work will attempt to associate with relevant parameters of the eruption. The models explored are for WD masses of 0.65 and 0.70 $M_\odot$ with constant accretion rates of $10^{-9}-10^{-11} M_\odot yr^{-1}$ and a core temperature ($\rm T_c$) of 30 or 50MK, and one self consistent binary with a WD mass of 0.45$M_\odot$ and an orbital period of $\sim3.8$ hours --- the orbital period derived for V1391 Cas by \cite{Schmidt2021}.

\begin{figure}
	\begin{center}
		{\includegraphics[trim={0.0cm 0.0cm 0.0cm 0.0cm},clip,	width=0.99\columnwidth]{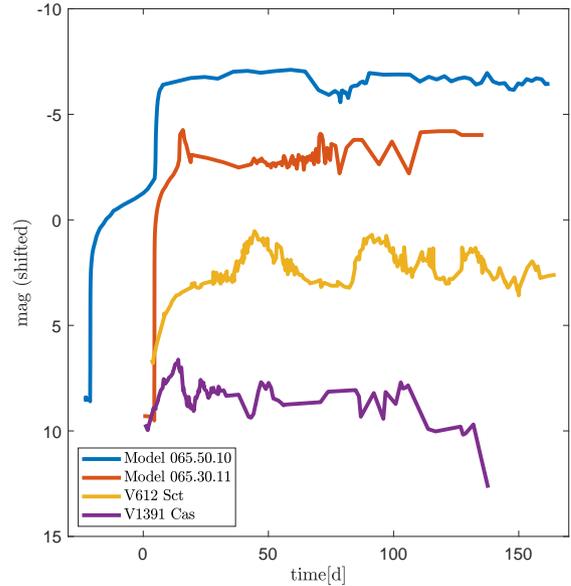}}
		\caption{Time from maximum for the early light curves of ASASSN-17hx (yellow) and V1391 Cas (purple) and two models (from Hillman et al. 2014a) with WD masses of 0.65$M_\odot$; one with $\rm T_c=50MK$ and $\dot{M}=10^{-10}M_\odot yr^{-1}$ (blue) and the other with $\rm T_c=30MK$ and $\dot{M}=10^{-11}M_\odot yr^{-1}$ (red). The magnitudes are shifted arbitrarily for convenience.}\label{fig:mags}
	\end{center}
\end{figure}

\section{Results}\label{sec:results}
Using the light curves from \cite{Hillman2014} as a guideline, two models were initially used for comparison with the behavior of V612 Sct and V1391 Cas --- the systems that were chosen here as examples of nova eruptions that feature prolonged fluctuations at maximum, as explained above. Figure \ref{fig:mags} shows the visual light curves of the two systems (reproduced from \cite{Poggiani2018} and \cite{Dubovsky2021} accordingly) as well as the visual brightness of two models from \cite{Hillman2014}, both of a 0.65$\rm M_\odot$ WD, one with an initial core temperature ($\rm T_c$) of 50MK and a constant accretion rate ($\dot{M}$) of $10^{-10}M_\odot yr^{-1}$ (model denoted 065.50.10) and the other with an initial $\rm T_c$ of 30MK and an $\dot{M}$ of $10^{-11}M_\odot yr^{-1}$ (model denoted 065.30.11). These two models bear a remarkable resemblance to the characteristics of the two chosen observed light curves --- a large amplitude, and oscillations of order ten days and 1-3 magnitudes that last a couple hundred years. Moreover, the lower $\dot{M}$ model (065.30.11, red curve in Figure \ref{fig:mags}) also shows a pre-maximum halt followed by a peak and dip just as seen in the V612 Sct light curve. 

\begin{figure}
	\begin{center}
		{\includegraphics[trim={0.0cm 1.8cm 0.0cm 1.0cm},clip,	width=.99\columnwidth]{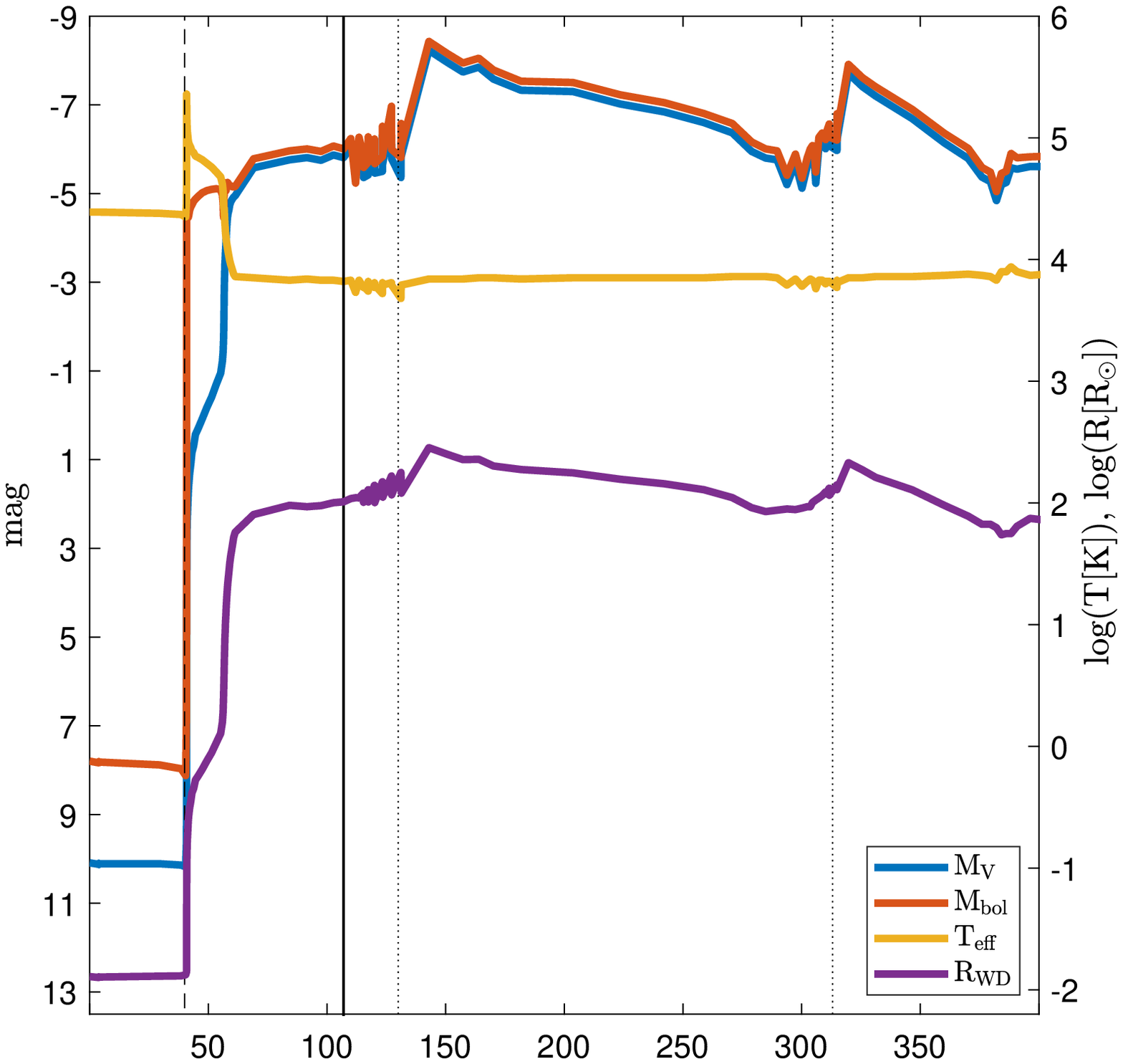}}
		{\includegraphics[trim={0.0cm 0.0cm 0.0cm 0.0cm},clip,	width=.99\columnwidth]{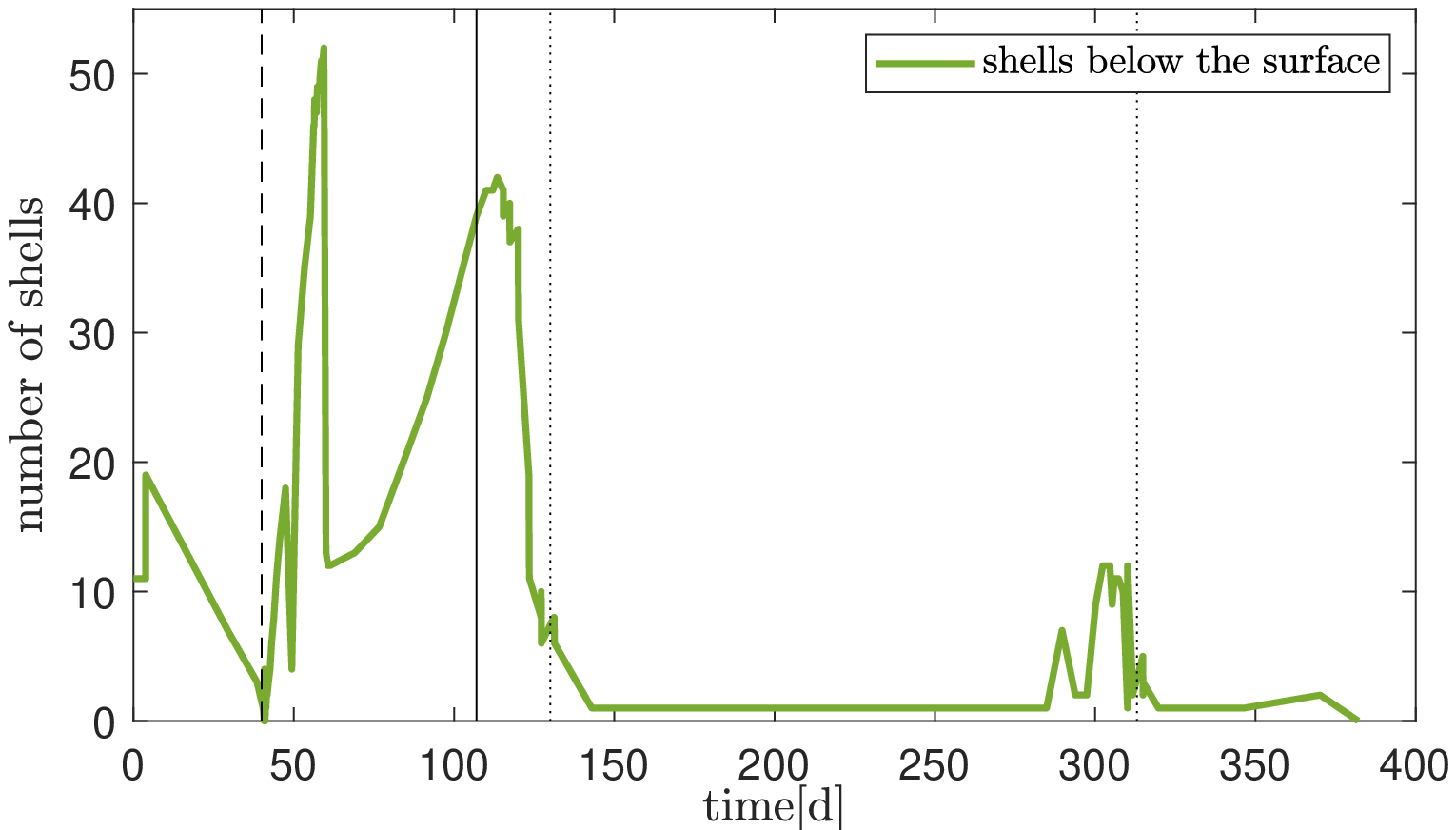}}
		\caption{Time from maximum for a model of a 0.65$M_\odot$ WD with an initial $\rm T_c=30MK$ and a constant $\dot{M}=10^{-10}M_\odot yr^{-1}$. Top: Visual magnitude (from Hillman et al. 2014a, blue); bolometric magnitude (red); $\rm T_{eff}$ (yellow) and WD radius (purple). Bottom: a qualitative measure of convection stability - number of shells below the surface where the convective front resides. The dashed black line marks the onset of the TNR, the solid black line marks the beginning of mass loss and the dotted black lines are for eye-guidance.}\label{fig:convec0653010}
	\end{center}
\end{figure}

It has been established that the WD mass, the accretion rate and the initial core temperature are the basic parameters that determine the outcome of a nova cycle, the core temperature being the least dominant of the three \cite[e.g.,][]{Kovetz1994,Schwartzman1994,Prikov1995,Yaron2005}. This study shows that the core temperature has an effect on the magnitude of the oscillations during the plateaued maximum. To demonstrate this, plotted in Figure \ref{fig:convec0653010}, is the light curve of a model with a WD mass of 0.65$M_\odot$ and an accretion rate of $10^{-10}M_\odot yr^{-1}$ --- the same as the blue curve in Figure \ref{fig:mags}, but with an initial core temperature of 30MK rather than 50MK (model denoted 065.30.10). The fluctuations here are on a longer time scale, and of larger amplitude than the model with the same accretion rate but higher temperature (model 065.50.10, blue curve in Figure \ref{fig:mags}), and than the model with the same temperature, but lower accretion rate (model 065.30.11, red curve in Figure \ref{fig:mags}). 

The large scale fluctuations exhibited in Figure \ref{fig:convec0653010} are used here as an advantage in investigating the cause us such a behavior. For this purpose, additionally plotted in Figure \ref{fig:convec0653010}, is the bolometric magnitude ($\rm M_{bol}$, red curve), being a couple magnitudes higher than visual at quiescence (to the left of the black dashed line), an even larger difference right at the onset of the TNR (the black dashed line), and then, the two curves become virtually coincidental. This is immediately justified by the changing effective temperature ($\rm T_{eff}$, yellow curve). At the onset of TNR the temperature rises steeply, and the expansion lags behind momentarily (purple curve representing the WD radius, $\rm R_{WD}$). For relatively low mass WDs the pressure exerted on the envelope is relatively low, resulting in a longer lag time. This is when pre-maximum UV flashes are predicted by models such as this one \cite[]{Hillman2014}, and are indeed observed in some actual nova as well \cite[]{Pietsch2007A,Pietsch2007B,Cao2012}. The envelope then expands, allowing the effective temperature to drop, causing the visual part of the spectrum to become more pronounced. For this model the two brightness curves (bolometric and visual) become nearly coincident during mass loss because the $\rm T_{eff}$ is of order $\sim 7000$K --- i.e., peaks in the visual, thus, most of the energy radiates in the visual regime. 

\begin{figure}
	\begin{center}
		{\includegraphics[trim={0.0cm 1.8cm 0.0cm 1.0cm},clip,	width=.99\columnwidth]{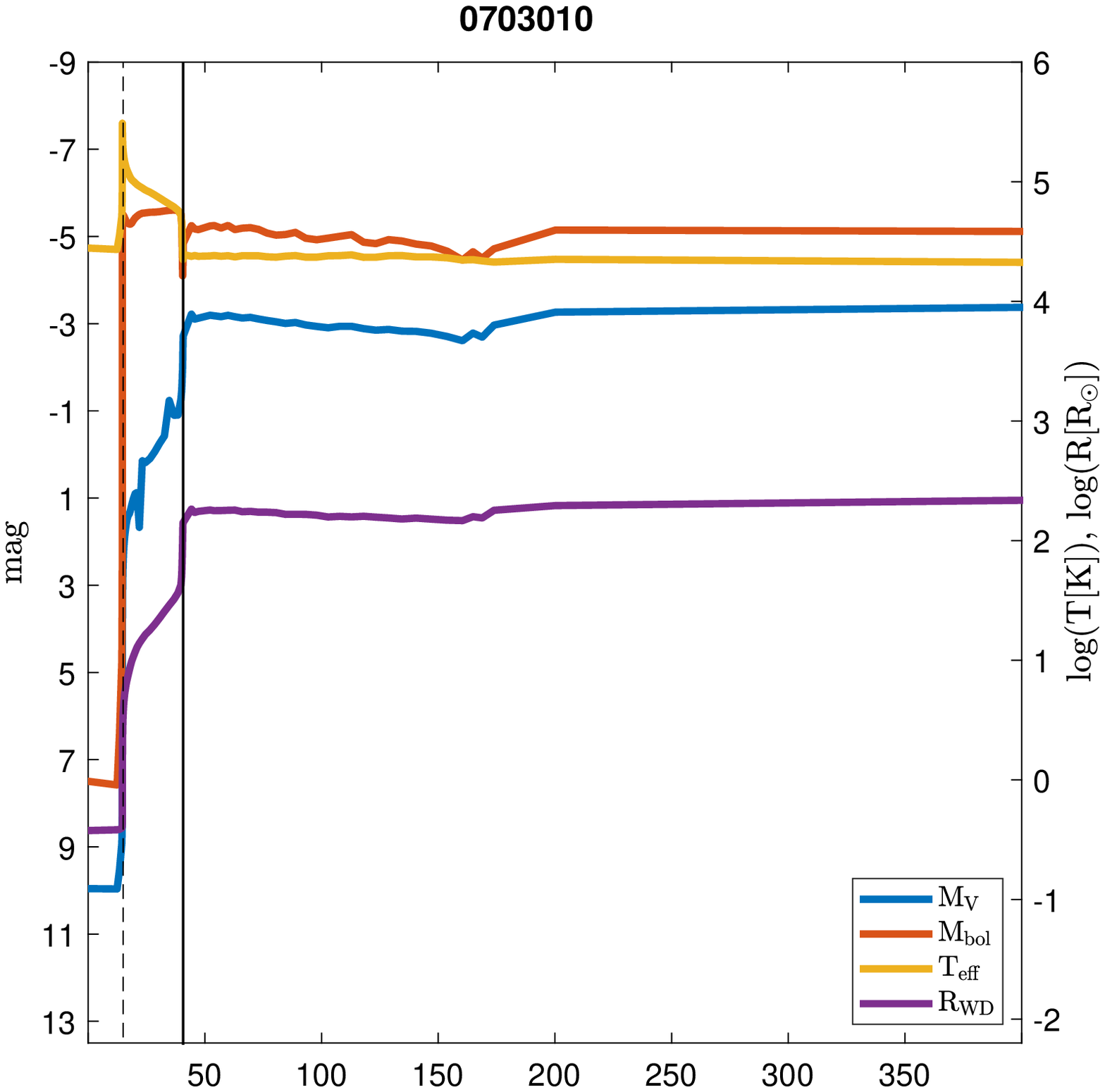}}
		
		{\includegraphics[trim={0.0cm 0.0cm 0.0cm 0.0cm},clip,	width=.99\columnwidth]{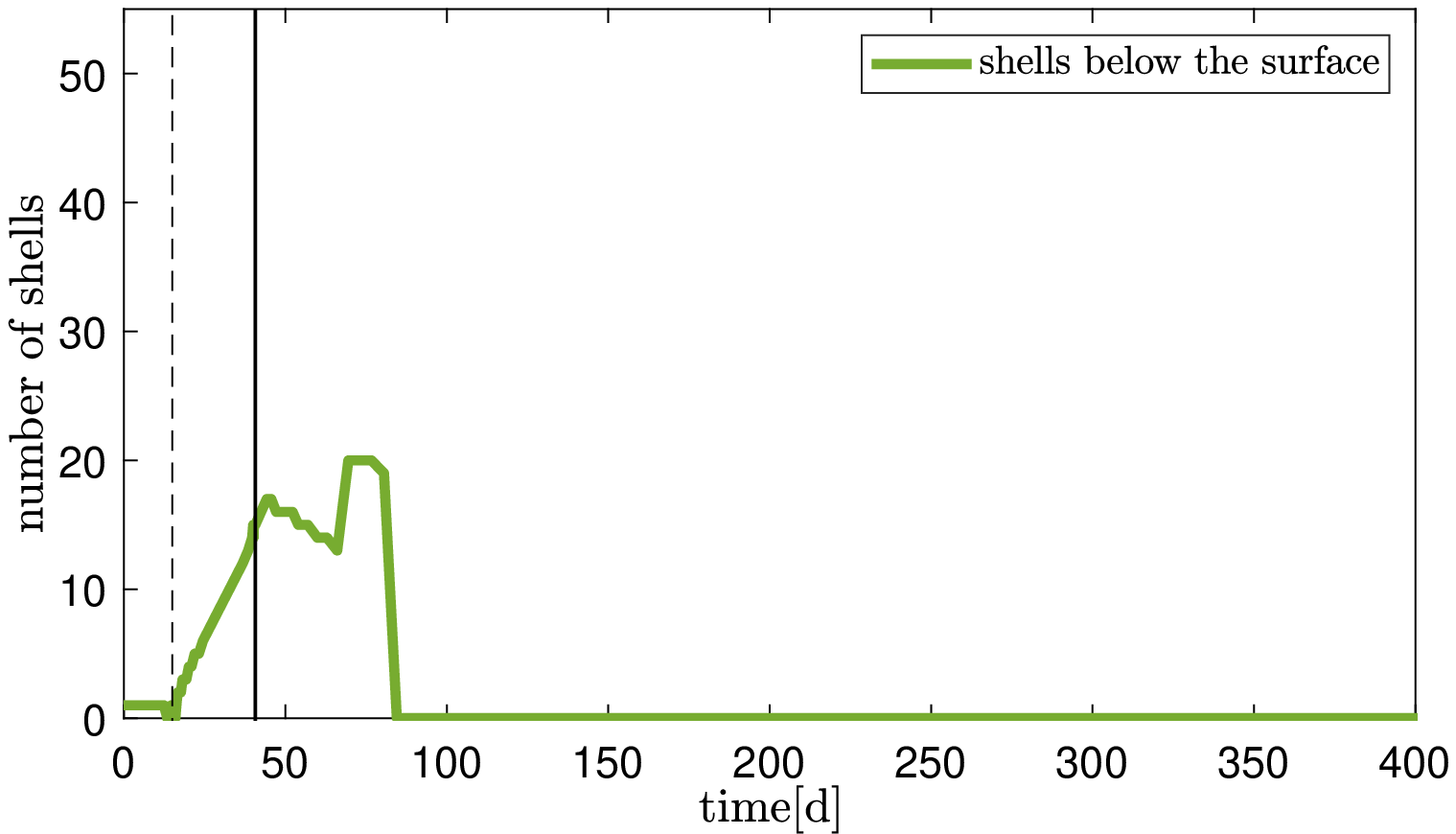}}
		\caption{Description as in Figure \ref{fig:convec0653010}, for a 0.7$M_\odot$ WD.} \label{fig:convec0703010}
	\end{center}
\end{figure}

In contrast, as another example, Figure \ref{fig:convec0703010} shows the behavior of a somewhat more massive WD --- 0.7$M_\odot$ with the same $\rm T_c$ and $\dot{M}$ (model denoted 070.30.10) as in Figure \ref{fig:convec0653010}, but revealing a much higher $\rm T_{eff}$, even during eruption, of order $\sim15000$K, placing the visual magnitude substantially lower than the bolometric the entire cycle. This is also in agreement with \cite{Sparks1978} who found for HR Del a bolometric magnitude that is higher than the visual. \cite{Shara2018} estimate the WD mass in HR Del to be $\sim0.84M_\odot$ explaining the magnitude difference in the same manner as the 0.7$M_\odot$ model here. In this such case, The UV band would be closer to the bolometric magnitude than the visual. The 0.7$M_\odot$ WD model is less than 10$\%$ more massive than the 0.65$M_\odot$ WD model, and the other parameters are identical, and yet it does not show the fluctuations that this work is focused on. 

Returning to Figure \ref{fig:convec0653010}, it is clear that the WD radius ($\rm R_{WD}$) follows the trend of the brightness, while $T\rm_{eff}$ remains stable, meaning that the fluctuations in the brightness are a direct result of a varying radius\footnote{Abiding by the Stefan-Boltzmann law: $\rm L_{WD,bol}\propto R^2_{WD}T^4_{eff}$} --- inflation and contraction. The comparison, 0.7$\rm M_\odot$ WD light curve in Figure \ref{fig:convec0703010}, which does not exhibit brightness fluctuations does not exhibit variations in the WD radius either. This points to the WD's radius, i.e., its envelope, as the driver of the light curve oscillations, leaving no need to look for an explanation in the expanding shell.

So what is causing the radius to fluctuate? The lower panels of Figures \ref{fig:convec0653010} and \ref{fig:convec0703010} demonstrate (qualitatively, via number of shells) how close the convection is to the surface, revealing large fluctuations that coincide with the brightness peaks in Figure \ref{fig:convec0653010}, whereas Figure \ref{fig:convec0703010} shows a slow increase and then decrease in the number of convective shells, but absent any sharp fluctuations. So this solves the mystery --- the oscillations seen in Figure \ref{fig:convec0653010}, and therefore may be deduced also for the two models in Figure \ref{fig:mags}, are a result of the convection front approaching the surface and receding from it repeatedly.


This conclusion immediately raises the followup question --- why are these fluctuations at a plateaued maximum seen in some models and not in others? 
\cite{Mason2020} explain that oscillations are favored in low mass WDs because their gravitational pull is weaker, allowing more time for the expanding envelope to respond to radiative diffusion. This is in addition to the envelope-to-WD mass ratio being orders of magnitude larger for low mass WDs, so mixing takes longer, meaning that the envelope will be less uniform for lower mass WDs.
The models here show that this mechanism is highly sensitive to the time scale of accretion, explaining why for the same WD mass (0.65$M_\odot$) and the same core temperature (30MK) the lower accretion rate --- i.e., a $\sim10$ times longer accretion phase --- brings the envelope to a more uniform state before erupting, thus the fluctuations are less pronounced. The higher WD mass manages to almost avoid them due to a smaller envelope.

\begin{figure}
	\begin{center}
		{\includegraphics[trim={0.7cm 0.0cm 0.9cm 0.0cm},clip,	width=.99\columnwidth]{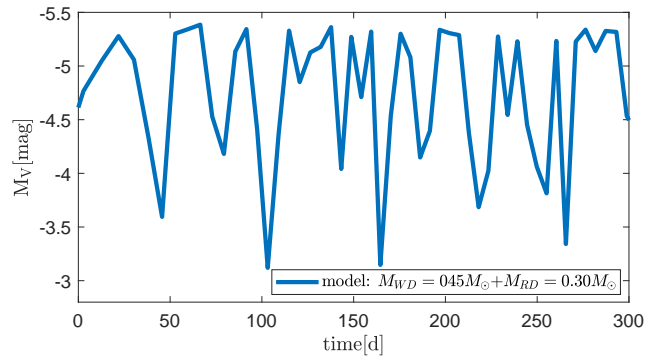}}
		\caption{Visual magnitude vs. time from maximum for a self consistent model with $M_{WD}=0.45M_\odot$ at the evolutionary point where $P_{orb}\sim3$hr.}\label{fig:045045zoom}
	\end{center}
\end{figure}

Following the conclusion that the WD mass for this type of behavior must be low, in order to produce these fluctuations, this work proceeds to the next model, of an extremely low mass WD of $0.45\rm M_\odot$ in order to attempt to obtain two observational characteristics. The first is an extremely large ejected mass, not typical of observed nova, and calculated for V612 Sct by \cite{Mason2020}. The second is an orbital period of order three hours, as deduced for V1391 Cas by \cite{Schmidt2021}. In this work, these characteristics were sought out via self consistent binary evolution \cite[]{Hillman2020,Hillman2021}, for a WD with a stellar mass of 0.45$\rm M_\odot$ at the evolutionary point where the companion mass is $\sim0.3\rm M_\odot$ i.e., has been eroded to below the magnetic braking limit\footnote{The donor mass was initially set at a more massive, 0.45$\rm M_\odot$ at first Roche lobe contact, and allowed to evolve freely.}. This choice of mass for the donor is in order to obtain roughly the required orbital period, which in CVs is roughly proportional to the RD's radius and mass \cite[]{Knigge2011a,Hillman2020,Hillman2020a}. The early light curve of this model is plotted in Figure \ref{fig:045045zoom}, showing fluctuations of order $\sim$2 magnitudes and tens of days --- as seen in V612 Sct and V1319 Cas. The ejected mass for this model is $\sim6\times10^{-4}M_\odot$ which is quite close to \cite{Mason2020}'s calculated estimate of 7 to 9$\times10^{-4}M_\odot$. 

\section{Discussion}\label{sec:discussion}
It has been shown in \cite{Hillman2020} and \cite{Hillman2021} that long term self consistent simulations of novae in Roche lobe overflowing CVs, show a stark difference in the average accretion rate between systems with fully convective donors, and systems with non-fully convective donors. The average accretion rate for the former being of order $10^{-11}M_\odot yr^{-1}$ and for the latter of order $10^{-9}M_\odot yr^{-1}$. The orbital period that seems to be typical of systems that show the fluctuations studied here is of order $\sim3-4$ hours \cite[]{Chochol2015,Schmidt2020,Schmidt2021} which in Roche lobe overflowing CVs corresponds to a RD mass of order $\sim0.3-0.4M_\odot$\footnote{This is because the orbital period at RLOF is roughly proportional to the RD mass via $M_{RD}\approx0.1P_{orb}$ \cite[e.g.,][]{Knigge2011a}.}. This stellar mass is right on the borderline of being fully convective \cite[]{Howell2001}. Since the magnetic braking of a star is inefficient for fully convective stars \cite[]{Howell2001}, for cases with such a donor mass, estimating the average accretion rate becomes problematic. This is because magnetic braking is the primary mechanism for removing angular momentum and pulling the stars closer together, which causes an increase in the RLOF and with it in the accretion rate. This is the primary reason that \cite{Hillman2021} has concluded that the average accretion rate for magnetic braking in full power is roughly two orders of magnitude higher than for total lack of magnetic braking. Since the estimated RD mass does not contribute to the knowing of whether or not the star is fully convective, it is not helpful in deducing an average accretion, which is the reason this work explored a wide range.

The results here agree with \cite{Mason2020} that insufficient time for mixing in an erupting massive envelope is a primary cause for the fluctuations. This may at first insinuate that the higher the accretion rate the better, because it will allow less mixing time, but a system with a high accretion rate ($\gtrapprox10^{-9}M_\odot yr^{-1}$) will also eject less mass, and since the accretion phase is short, it is poorly enriched, so more uniform to start with. The results here show that the development and magnitude of these fluctuations are sensitive not only to the accretion time and envelope mass, but also to the temperature underlying the accreted shell, and this is because the temperature plays a role in driving the convection.

\cite{Strope2010} have identified an entire class of nova eruptions that oscillate at maximum light. 	Their fig.13 shows this explicitly for V603 Aql, V868 Cen, V1494 Aql, GK Per, and V2467 Cyg, showing oscillations of order 1 mag for 1-20 days lasting 100 days. However, the estimated masses of the WDs in these system are of order $\sim1.0-1.2M_\odot$ and of the average accretion rate of order $\sim10^{-10}-10^{-9}M_\odot yr^{-1}$ \cite[]{Shara2018}. The magnitude of the fluctuations for these systems is on a somewhat smaller scale than exhibited by the 0.65$M_\odot$ model (with a similar accretion rate), which is smaller than those exhibited by the 0.45$M_\odot$ model, supporting the conclusion that as the WD mass is higher, the mechanisms at work reduce the fluctuations and even the possibility of them forming. 

So why does it seem there are more of these at higher WD masses than at lower? This is simply since there ARE more erupting novae with higher WD masses. Statistically speaking, the median observed (corrected for observational bias) WD mass in novae is $\sim1.13M_\odot$ ($\sim1.06M_\odot$) \cite[]{Shara2018, Hillman2020a}, and to top off the observational bias, higher mass WDs also erupt more often, leading to more detections of nova with high WD masses. 

\section{Conclusions}\label{sec:conclusions}
This work explored a handful of models, including one self consistent, in order to investigate the possibility of a prolonged and fluctuating maximum being caused internally, by the WD itself rather than as a phenomenon occurring in the ejected, expanding shell. The results show that the changes in the convection are the cause of these fluctuations, and that these changes are a result of insufficient time for mixing the envelope, causing the non-uniformity to lead to a competition between convective mixing and radial inflation, resulting in the convective front approaching and receding from the surface periodically. 

This will occur for the right combination of accretion time, WD mass and the temperature at the base of the accreted envelope, and this combination will determine the size and time scale of the fluctuation. 

The low mass WDs are strong candidates for forming them since they have massive envelopes and a large envelope-to-WD-mass ratio, requiring more time for mixing. This is why (for a given $\rm M_{WD}$ and $\rm T_c$) the higher accretion rate model ($10^{-10}M_\odot yr^{-1}$), i.e., shorter accretion time, shows larger fluctuations than the lower accretion rate model ($10^{-10}M_\odot yr^{-1}$), i.e., longer accretion time. The parameters that are suitable for producing these fluctuations, are the same that are expected to produce long PMHs, since both phenomena require a long accretion period.

Although this worked shows that these fluctuations can be explained as originating from the WD itself, the possibility of observing similar features originating in an expanding ejected shell or the donor, such as suggested by \cite{Aydi2020} and \cite{Dubovsky2021} still remains, and may possibly explain their presence, if detected, in WDs with high masses and/or accretion rates. 
\section*{Acknowledgements}
The support from the Authority for Research $\&$ Development and
the chairman of the Department of Physics in Ariel University are
gratefully acknowledged.
\section*{Data availability}
The data underlying this article will be shared on reasonable request to the author.

\bibliographystyle{aasjournal}
\bibliography{rfrncs}
\end{document}